# FUTURE-AI: International consensus guideline for trustworthy and deployable artificial intelligence in healthcare

Karim Lekadir, PhD,[1,2*], Aasa Feragen, PhD,[3], Abdul Joseph Fofanah, MEng,[4], Alejandro F Frangi, PhD,[5,6], Alena Buyx, PhD,[7], Anais Emelie, MEng,[1], Andrea Lara, PhD,[8], Antonio R Porras, PhD,[9], An-Wen Chan, DPhil,[10], Arcadi Navarro, PhD,[2,11], Ben Glocker, PhD,[12], Benard O Botwe, PhD,[13,14], Bishesh Khanal, PhD,[15], Brigit Beger, MA,[16], Carol C Wu, MD,[17], Celia Cintas, PhD,[18], Curtis P Langlotz, MD,[19], Daniel Rueckert, PhD,[20,21], Deogratias Mzurikwao, PhD,[22], Dimitrios I Fotiadis, PhD,[23], Doszhan Zhussupov,[24], Enzo Ferrante, PhD,[25], Erik Meijering, PhD,[26], Eva Weicken, MD,[27], Fabio A González, PhD,[28], Folkert W Asselbergs, MD,[29,30], Fred Prior, PhD,[31], Gabriel P Krestin, MD,[32], Gary S Collins, PhD,[33], Geletaw S Tegenaw, PhD,[34], Georgios Kaissis, PhD,[20], Gianluca Misuraca, PhD,[35], Gianna Tsakou, Dip,[36], Girish Dwivedi, MD,[37], Haridimos Kondylakis, PhD,[38], Harsha Jayakody, MD,[39], Henry C Woodruf, PhD,[40], Horst Joachim Mayer, PhD,[41], Hugo JWL Aerts, PhD,[42], Ian Walsh, PhD,[43], Ioanna Chouvarda, PhD,[44], Irène Buvat, PhD,[45], Isabell Tributsch, MA, [1], Islem Rekik, PhD,[46,47], James Duncan, PhD,[48], Jayashree Kalpathy-Cramer, PhD,[49], Jihad Zahir, PhD,[50], Jinah Park, PhD,[51], John Mongan, MD,[52], Judy W Gichoya, MD,[53], Julia A Schnabel, PhD,[54], Kaisar Kushibar, PhD,[1], Katrine Riklund, MD,[55], Kensaku Mori, PhD,[56], Kostas Marias, PhD,[38], Lameck M Amugongo, PhD,[57], Lauren A Fromont, PhD,[58], Lena Maier-Hein, PhD,[59], Leonor Cerdá Alberich, PhD,[60], Leticia Rittner, PhD,[61], Lighton Phiri, PhD,[62], Linda Marrakchi-Kacem, PhD,[63], Lluís Donoso-Bach, MD,[64], Luis Martí-Bonmatí, MD,[65], M Jorge Cardoso, PhD,[66], Maciej Bobowicz, MD,[67], Mahsa Shabani, PhD,[68], Manolis Tsiknakis, PhD,[38], Maria A Zuluaga, PhD,[69], Maria Bielikova, PhD,[70], Marie-Christine Fritzsche,[7], Marina Camacho, MSc,[1], Marius George Linguraru, DPhil,[71], Markus Wenzel, PhD,[27], Marleen De Bruijne, PhD,[32], Martin G Tolsgaard, MD,[72], Marzyeh Ghassemi, PhD,[73], Md Ashrafuzzaman, PhD,[74], Melanie Goisauf, PhD,[75], Mohammad Yaqub, PhD,[76], Mónica Cano Abadía, PhD,[75], Mukhtar M E Mahmoud, PhD,[77], Mustafa Elattar, PhD,[78], Nicola Rieke, PhD,[79], Nikolaos Papanikolaou, PhD,[80], Noussair Lazrak, PhD,[1], Oliver Díaz, PhD,[1], Olivier Salvado, PhD,[81], Oriol Pujol, PhD,[1], Ousmane Sall, PhD,[82], Pamela Guevara, PhD,[83], Peter Gordebeke, MSc,[84], Philippe Lambin, MD,[40], Pieta Brown, MSc,[85], Purang Abolmaesumi, PhD,[86], Qi Dou, PhD,[87], Qinghua Lu, PhD,[81], Richard Osuala, MSc,[1], Rose Nakasi, PhD,[88], S Kevin Zhou, PhD,[89], Sandy Napel, PhD,[90], Sara Colantonio, PhD,[91], Shadi Albarqouni, PhD,[92], Smriti Joshi, MSc,[1], Stacy Carter, PhD,[93],

Stefan Klein, PhD,[32], Steffen E Petersen, PhD,[94], Susanna Aussó, MSc,[95], Suyash Awate, PhD,[96], Tammy Riklin Raviv, PhD,[97], Tessa Cook, MD,[98], Tinashe E M Mutsvangwa, PhD,[99], Wendy A Rogers, PhD,[100], Wiro J Niessen, PhD,[32], Xènia Puig-Bosch, PhD,[1], Yi Zeng, PhD,[101], Yunusa G Mohammed, PhD,[102], Yves Saint James Aquino, PhD,[93], Zohaib Salahuddin, MSc,[40], Martijn P A Starmans, PhD,[32].

**AFFILIATIONS:**

[1] Artificial Intelligence in Medicine Lab (BCN-AIM), Department de Matemàtiques i Informàtica, Universitat de Barcelona, Barcelona, Spain

[2] Institució Catalana de Recerca i Estudis Avançats (ICREA), Barcelona, Spain

[3] DTU Compute, Technical University of Denmark, Kgs Lyngby, Denmark

[4] Department of Mathematics and Computer Science, Faculty of Science and Technology, Milton Margai Technical University, Freetown, Sierra Leone

[5] Center for Computational Imaging & Simulation Technologies in Biomedicine, Schools of Computing and Medicine, University of Leeds, Leeds, United Kingdom

[6] Medical Imaging Research Center (MIRC), Cardiovascular Science and Electronic Engineering Departments, KU Leuven, Leuven, Belgium

[7] Institute of History and Ethics in Medicine, Technical University of Munich, Munich, Germany

[8] Faculty of Engineering of Systems, Informatics and Sciences of Computing, Galileo University, Guatemala City, Guatemala

[9] Department of Biostatistics and Informatics, Colorado School of Public Health, University of Colorado Anschutz Medical Campus, Aurora, CO, United States

[10] Department of Medicine, Women's College Research Institute, University of Toronto, Toronto, Canada

[11] Universitat Pompeu Fabra and BarcelonaBeta Brain Research Center, Barcelona, Spain

[12] Department of Computing, Imperial College London, London, United Kingdom

[13] School of Biomedical & Allied Health Sciences, University of Ghana, Accra, Ghana

[14] Department of Midwifery & Radiography, School of Health & Psychological Sciences, City University of London, UK

[15] NepAl Applied Mathematics and Informatics Institute for research (NAAMII), Kathmandu, Nepal

[16] European Heart Network, Brussels, Belgium

[17] Department of Thoracic Imaging, University of Texas MD Anderson Cancer Center, Houston, United States

[18] IBM Research Africa, Nairobi, Kenya

[19] Departments of Radiology, Medicine, and Biomedical Data Science, Stanford University School of Medicine, Stanford, United States

[20] Institute for AI and Informatics in Medicine, Klinikum rechts der Isar, Technical University Munich, Munich, Germany

[21] Department of Computing, Imperial College London, London, UK

[22] Muhimbili University of Health and Allied Sciences, Dar es Salaam, Tanzania

[23] Unit of Medical Technology and Intelligent Information Systems, Foundation for Research and Technology - Hellas (FORTH), Ioannina, Greece

[24] Almaty AI Lab, Almaty, Kazakhstan

[25] CONICET, Universidad Nacional del Litoral, Santa Fe, Argentina

[26] School of Computer Science and Engineering, University of New South Wales, Sydney, Australia

[27] Fraunhofer Heinrich Hertz Institute, Berlin, Germany

[28] Computing Systems and Industrial Engineering Dept., Universidad Nacional de Colombia, Bogotá, Colombia

[29] Amsterdam University Medical Centers, Department of Cardiology, University of Amsterdam, Amsterdam, The Netherlands

[30] Health Data Research UK and Institute of Health Informatics, University College London, London, United Kingdom

[31] Department of Biomedical Informatics, University of Arkansas for Medical Sciences, Little Rock, United States

[32] Department of Radiology & Nuclear Medicine, Erasmus MC University Medical Center, Rotterdam, the Netherlands

[33] Centre for Statistics in Medicine, University of Oxford, Oxford, United Kingdom

[34] Faculty of Computing and Informatics, JiT, Jimma University, Jimma, Ethiopia

[35] Department of Artificial Intelligence, Universidad Politécnica de Madrid, Madrid, Spain

[36] Gruppo Maggioli, Research and Development Lab, Athens, Greece

[37] Department of Advanced Clinical and Translational Cardiovascular Imaging, The University of Western Australia, Perth, Australia
[38] Department of Electrical and Computer Engineering, Hellenic Mediterranean University and Foundation for Research and Technology - Hellas (FORTH), Crete, Greece
[39] Postgraduate Institute of Medicine, University of Colombo, Colombo, Sri Lanka
[40] The D-lab, Department of Precision Medicine, GROW - School for Oncology and Reproduction, Maastricht University, Maastricht, the Netherlands
[41] CE Plus, a division of regenold GmbH, Badenweiler, Germany
[42] Artificial Intelligence in Medicine Program, Mass General Brigham, Harvard Medical School, Boston, United States
[43] Bioprocessing Technology Institute, Agency for Science, Technology and Research (A*STAR), Singapore, Singapore
[44] School of Medicine, Aristotle University of Thessaloniki, Thessaloniki, Greece
[45] Institut Curie, Inserm, Orsay, France
[46] BASIRA Lab, Imperial-X and Computing Department, Imperial College London, UK
[47] Faculty of Computer and Informatics Engineering, Istanbul Technical University, Turkey
[48] Departments of Biomedical Engineering and Radiology & Biomedical Imaging, Schools of Engineering & Applied Science and Medicine, Yale University, New Haven, United States
[49] Massachusetts General Hospital, Harvard Medical School, Massachusetts, United States
[50] LISI Laboratory, Computer Science Department, Cadi Ayyad University, Marrakech, Morocco
[51] School of Computing, Korea Advanced Institute of Science and Technology, Daejeon, South Korea
[52] Department of Radiology and Biomedical Imaging, University of California San Francisco, San Francisco, United States
[53] Department of Radiology & Imaging Sciences, Emory University, Atlanta, United States
[54] Institute of Machine Learning in Biomedical Imaging, Helmholtz Center Munich, Munich, Germany
[55] Department of Radiation Sciences, Diagnostic Radiology, Umeå University, Umeå, Sweden
[56] Graduate School of Informatics, Nagoya University, Nagoya, Japan
[57] Department of Software Engineering, Namibia University of Science & Technology, Windhoek, Namibia

[58] Center for Genomic Regulation, The Barcelona Institute of Science and Technology, Barcelona, Spain
[59] Div. Intelligent Medical Systems (IMSY), German Cancer Research Center (DKFZ), Heidelberg, Germany
[60] Biomedical Imaging Research Group, La Fe Health Research Institute, Valencia, Spain
[61] School of Electrical and Computer Engineering, University of Campinas, Campinas, Brazil
[62] Department of Library Information Science, University of Zambia, Lusaka, Zambia
[63] National Engineering School of Tunis, University of Tunis El Manar, Tunis, Tunisia
[64] Clinical Advanced Technologies institute (CATI), Hospital Clínic of Barcelona, Barcelona, Spain
[65] Medical Imaging Department, Hospital Universitario y Politécnico La Fe, Valencia, Spain
[66] School of Biomedical Engineering & Imaging Sciences, King's College London, London, United Kingdom
[67] 2nd Division of Radiology, Medical University of Gdansk, Gdansk, Poland
[68] Faculty of Law and Criminology, Ghent University, Ghent, Belgium
[69] Data Science Department, EURECOM, Sophia Antipolis, France
[70] Kempelen Institute of Intelligent Technologies, Bratislava, Slovakia
[71] Sheikh Zayed Institute for Pediatric Surgical Innovation, Children’s National Hospital Washington DC, United States
[72] Copenhagen Academy for Medical Education and Simulation Rigshospitalet, University of, Copenhagen, Copenhagen, Denmark
[73] Electrical Engineering and Computer Science (EECS) and Institute for Medical Engineering & Science (IMES), Massachusetts Institute of Technology, Cambridge, United States
[74] Department of Biomedical Engineering, Military Institute of Science and Technology, Dhaka, Bangladesh
[75] BBMRI-ERIC, ELSI Services & Research, Graz, Austria
[76] Mohamed Bin Zayed University of Artificial Intelligence, Abu Dhabi, United Arab Emirates
[77] Faculty of Computer Science and Information Technology, University of Kassala, Kassala, Sudan
[78] Center for Informatics Science, Nile University, Sheikh Zayed City, Egypt
[79] Healthcare & Life Science EMEA, NVIDIA GmbH, Munich, Germany

[80] Computational Clinical Imaging Group, Champalimaud Foundation, Lisbon, Portugal
[81] Data61, The Commonwealth Scientific and Industrial Research Organisation (CSIRO), Canberra, Australia
[82] Pôle Sciences, Technologies et Numérique, Université Virtuelle du Sénégal, Diamniadio, Senegal
[83] Faculty of Engineering, Universidad de Concepción, Concepción, Chile
[84] European Institute for Biomedical Imaging Research, Vienna, Austria
[85] Orion Health, Auckland, New Zealand
[86] Department of Electrical and Computer Engineering, the University of British Columbia, Vancouver, BC, Canada
[87] Department of Computer Science and Engineering, The Chinese University of Hong Kong, Hong Kong, China
[88] Makerere Artificial Intelligence Lab, Makerere University, Kampala, Uganda
[89] School of Biomedical Engineering & Suzhou Institute for Advanced Research, University of Science and Technology of China, Suzhou, China
[90] Integrative Biomedical Imaging Informatics at Stanford (IBIIS), Department of Radiology, Stanford University, Stanford CA, United States
[91] Institute of Information Science and Technologies of the National Research Council of Italy, Pisa, Italy
[92] Department of Diagnostic and Interventional Radiology, University Hospital Bonn, Bonn, Germany
[93] Australian Centre for Health Engagement, Evidence and Values, School of Health and Society, University of Wollongong, New South Wales, Australia
[94] William Harvey Research Institute, Queen Mary University of London, London, United Kingdom
[95] Artificial Intelligence in Healthcare Program, TIC Salut Social Foundation, Barcelona, Spain
[96] Computer Science and Engineering Department, Indian Institute of Technology Bombay, Mumbai, India
[97] The School of Electrical and Computer Engineering, Ben-Gurion University, Beer Sheba, Israel
[98] Department of Radiology, Perelman School of Medicine at the University of Pennsylvania, Philadelphia, PA, United States

[99] Department of Human Biology, University of Cape Town, Cape Town, South Africa

[100] Department of Philosophy, and School of Medicine, Macquarie University, Sydney, Australia

[101] Center for Long-term AI, Chinese Academy of Sciences, Beijing, China

[102] Department of Human Anatomy, Gombe State University, Gombe, Nigeria

[*] **Corresponding author**: Karim Lekadir, PhD. E-mail: karim.lekadir@ub.edu

Artificial Intelligence in Medicine Lab (BCN-AIM)

Department de Matemàtiques i Informàtica

Universitat de Barcelona

Gran Via de les Corts Catalanes, 585, 08007 Barcelona, Spain

## ABSTRACT

**Background:** Despite major advances in artificial intelligence (AI) research for healthcare, the deployment and adoption of AI technologies remain limited in clinical practice. In recent years, concerns have been raised about the clinical, technical, ethical and legal risks associated with healthcare AI. To increase adoption in the real world, it is essential that AI technologies are trusted and accepted by patients, clinicians, health organisations and authorities. This paper describes the FUTURE-AI framework as the first international consensus guideline for trustworthy AI in healthcare.

**Methods:** The FUTURE-AI consortium was founded in 2021 and now comprises 117 interdisciplinary experts from 50 countries representing all continents, including AI scientists, clinical researchers, biomedical ethicists, and social scientists. Over a two-year period, the consortium established guiding principles and best practices for trustworthy and deployable AI through an iterative process comprising an in-depth literature review, a modified Delphi survey, and online consensus meetings.

**Findings:** The FUTURE-AI framework was established based on six guiding principles for trustworthy AI in healthcare, *i.e.* Fairness, Universality, Traceability, Usability, Robustness and Explainability. Through consensus, a set of 30 best practices were defined, addressing technical, clinical, socio-ethical and legal dimensions of trustworthy AI. The recommendations cover the entire lifecycle of healthcare AI, from design, development and validation to regulation, deployment, and monitoring.

**Interpretation:** FUTURE-AI is a structured, risk-informed framework which provides guidance for constructing healthcare AI tools that will be trusted, deployed and adopted in real-world clinical practice. Researchers are encouraged to take the recommendations into account in proof-of-concept stages to facilitate future translation towards clinical practice of healthcare AI.

## INTRODUCTION

Despite major advances in the field of healthcare AI, the deployment and adoption of AI technologies remain limited in real-world clinical practice. In recent years, concerns have been raised about the technical, clinical, ethical and societal risks associated with healthcare AI (1,2). In particular, existing research has shown that AI tools in healthcare can be prone to errors and patient harm, biases and increased health inequalities, lack of transparency and accountability, as well as data privacy and security breaches (3–7).

To increase adoption in the real world, it is essential that AI tools are trusted and accepted by patients, clinicians, health organisations and authorities. However, there is an absence of clear, widely accepted guidelines on how healthcare AI tools should be designed, developed, evaluated and deployed to be trustworthy, *i.e.* technically robust, clinically safe, ethically sound and legally compliant. To have a real impact at scale, such guidelines for responsible and trustworthy AI must be obtained through wide consensus involving international and inter-disciplinary experts.

In other domains, international consensus guidelines have made lasting impacts. For example, the FAIR guideline (8) for data management has been widely adopted by researchers, organisations and authorities, as they provide a structured framework for standardising and enhancing the tasks of data collection, curation, organisation and storage. While it can be argued that the FAIR principles do not cover every aspect of data management, as they focus more on findability, accessibility, interoperability and reusability of the data, and less on privacy and security, they delivered a code of practice that is now widely accepted and applied.

For AI in healthcare, initial efforts have focused on providing recommendations for the reporting of AI studies for different medical domains or clinical tasks (*e.g.* TRIPOD+AI (9), CLAIM(10), CONSORT-AI (11), DECIDE-AI (12), PROBAST-AI (13), CLEAR (14)). These guidelines do not provide best practices for the actual development and deployment of the AI tools but promote standardised and complete reporting of their development and evaluation. Recently, several researchers have published promising ideas on possible best practices for healthcare AI (15–22).

However, these proposals have not been established through wide international consensus and do not cover the whole lifecycle of healthcare AI (*i.e.* from design, development and validation to deployment, usage and monitoring).

In other initiatives, the World Health Organisation published a report focused on key ethical and legal challenges and considerations. As it was intended for health ministries and governmental agencies, it did not explore the technical and clinical aspects of trustworthy AI (23). Likewise, Europe's High-Level Expert Group on Artificial Intelligence established a comprehensive self-assessment checklist for AI developers. However, it covered AI in general and did not address the unique risks and challenges of AI in medicine and healthcare (24).

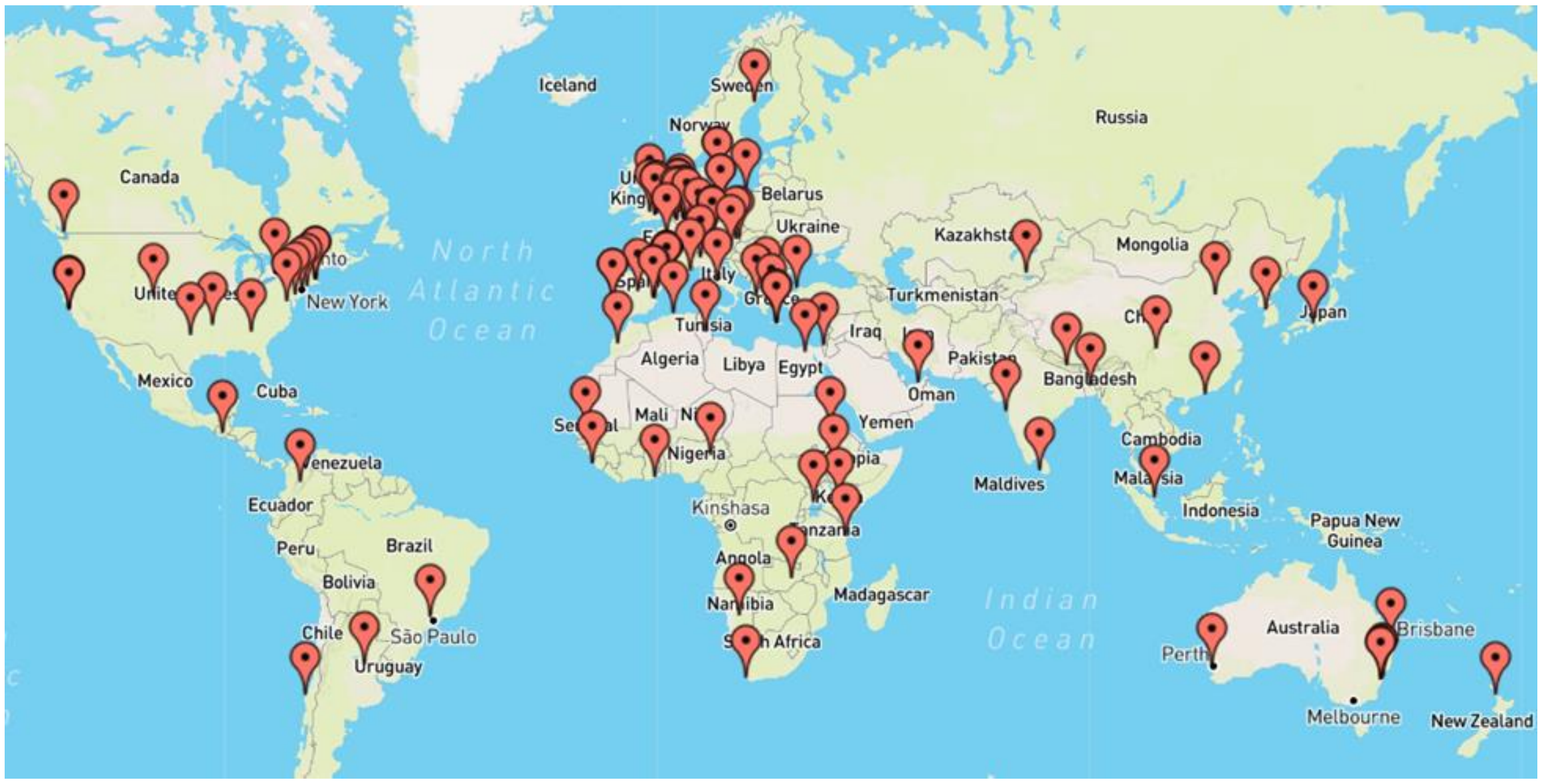

*Figure 1 – Geographical distribution of the multi-disciplinary experts.*

This paper addresses an important gap in the field of healthcare AI, by delivering the first structured and holistic guideline for trustworthy and ethical AI in healthcare, established through wide international consensus and covering the entire lifecycle of AI. The FUTURE-AI consortium was initiated in 2021 and currently comprises 117 international and inter-disciplinary experts from 50 countries (Figure 1), representing all continents (Europe, North America, South America, Asia, Africa, and Oceania). Additionally, the members represent a variety of disciplines (*e.g.* data science, medical research, clinical medicine, computer engineering, medical ethics, social sciences) and data domains (*e.g.* radiology, genomics, mobile health, electronic health records, surgery, pathology). To develop the FUTURE-AI framework, we drew inspiration from the FAIR principles for data management, and defined concise recommendations organised according to six guiding principles, *i.e.* Fairness, Universality, Traceability, Usability, Robustness and Explainability (Figure 2).

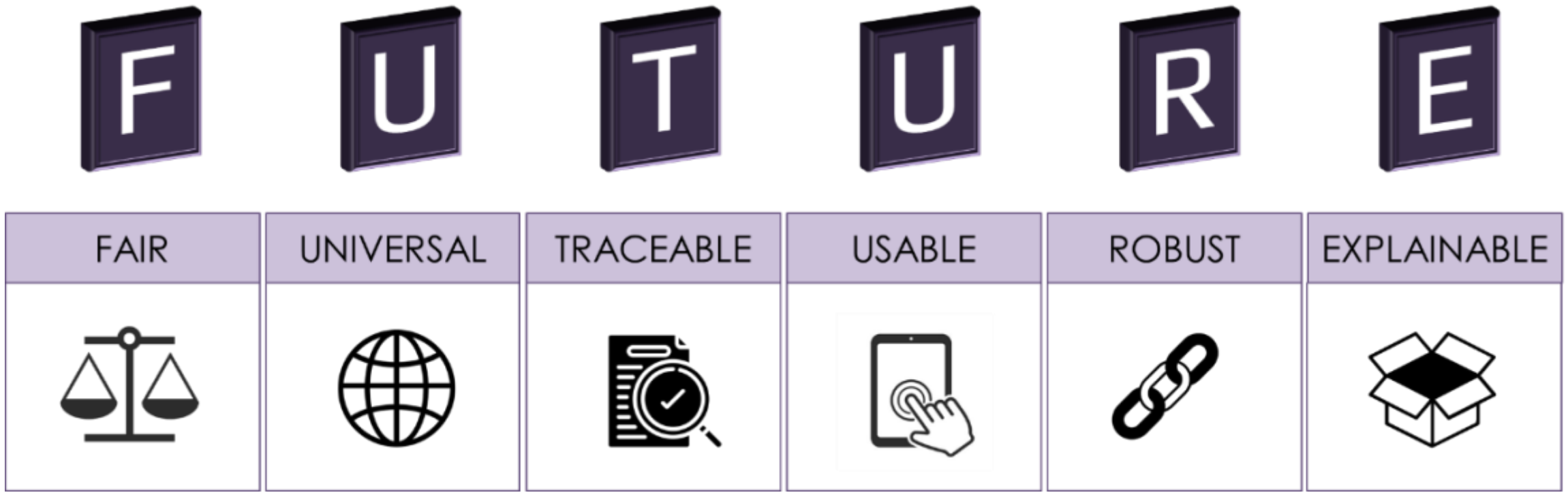

*Figure 2 – Organisation of the FUTURE-AI framework for trustworthy AI according to six guiding principles, i.e. Fairness, Universality, Traceability, Usability, Robustness and Explainability.*

## METHODS

FUTURE-AI is a structured framework that provides guiding principles as well as step-by-step recommendations for operationalising trustworthy and ethical AI in healthcare. This guideline was established through international consensus over a 24-month period using a modified Delphi approach (25,26). The process began with the definition of the six core guiding principles, followed by an initial set of recommendations, which was then subjected to seven rounds of extensive feedback and iterative discussions aimed at reaching consensus. In each round, we employed two complementary methods to aggregate the results: (1) a quantitative approach, which involved analysing the voting patterns of the experts to identify areas of consensus and disagreement; and (2) a qualitative approach, focusing on the synthesis of feedback and discussions based on recurring themes or new insights raised by multiple experts.

*Definition of the FUTURE-AI guiding principles:*

To develop a user-friendly guideline for trustworthy AI in medicine, we used the same approach as in the FAIR guideline, based upon a minimal set of guiding principles. Defining overarching guiding principles facilitates streamlining and structuring of best practices, as well as implementation by future end-users of the FUTURE-AI guideline.

To this end, we first reviewed the existing literature in healthcare AI, with a focus on the topics of trustworthy, responsible and ethical AI. This review enabled us to identify a wide range of requirements and dimensions often cited as essential for trustworthy AI. As shown in Table 1, these requirements were then thematically grouped, leading to our definition of the six core principles (*i.e.* Fairness, Universality, Traceability, Usability, Robustness and Explainability), which were arranged to form an easy-to-remember acronym (FUTURE-AI).

*Table 1 – Clustering of trustworthy AI requirements and selection of FUTURE-AI guiding principles.*

| | Clusters of requirements | Core principles |
|---|---|---|
| 1 | Fairness, Diversity, Inclusiveness, Non-discrimination, Unbiased AI, Equity | Fairness |
| 2 | Generalisability, Adaptability, Interoperability, Applicability, Universality | Universality |
| 3 | Traceability, Monitoring, Continuous learning, Auditing, Accountability | Traceability |
| 4 | Human-centred AI, User engagement, Usability, Accessibility, Efficiency | Usability |
| 5 | Robustness, Reliability, Resilience, Safety, Security | Robustness |
| 6 | Transparency, Explainability, Interpretability, Understandability | Explainability |

*Round 1. Definition of an initial set of recommendations:*

Six working groups composed of three experts each (including clinicians, data scientists and computer engineers) were created to explore the six guiding principles separately. The experts were recruited from five European projects (EuCanImage, ProCAncer-I, CHAIMELEON, PRIMAGE, INCISIVE), which together formed the AI for Health Imaging (AI4HI) network. By using "AI for medical imaging" as a common use case, each working group conducted a thorough literature review, then proposed a definition of the guiding principle in question, together with an initial list of best practices (between 6 and 10 for each guiding principle).

Subsequently, the working groups engaged in an iterative process of refining these preliminary recommendations via online meetings as well as by e-mail exchanges. At this stage, a degree of overlap and redundancy was identified across recommendations. For example, a recommendation to report any identified bias was initially proposed under both the Fairness and Traceability principles, while a recommendation to train the AI models with representative datasets appeared under Fairness and Robustness. After removing the redundancies and refining the formulations, a set of 55 preliminary recommendations was derived and then distributed to a broader panel of experts for further assessment, discussion, and refinement in the next round.

*Round 2. Online survey:*

In this round, the FUTURE-AI consortium was expanded to 72 members, by inviting new experts including AI scientists, healthcare practitioners, ethicists, social scientists, legal experts and industry professionals. The majority of the experts were recruited to complement the original consortium based on academic credentials, geographic location, and expertise. We then conducted an online survey to enable the experts to assess each recommendation using five options (Absolutely essential, Very important, Of average importance, Of little importance, Not important at all). The participants were also able to rate the formulation of the recommendation ("I would keep it as it is", "I would refine its definition") and propose modifications. Furthermore, they were able to propose merging recommendations or adding new ones. The survey included a section for free-text feedback on the core principles and the overall FUTURE-AI guideline.

The survey responses were quantitatively analysed to assess the consensus level. Recommendations that garnered a high-level agreement were selected for further discussion (>90%). On the other hand, recommendations that attracted significant negative feedback, which were particularly those that suggested specific methods over general guidelines, were discarded. The written feedback also prompted the merging of some recommendations, aiming to craft a more concise guideline for easier adoption by future users. Consequently, a revised list of 22 recommendations was derived, along with the identification of 16 contentious points for further discussions.

As part of the survey, we also sought feedback from the experts on the adequacy of these guiding principles in capturing the diverse requirements for trustworthy AI in healthcare. While the consensus among experts was largely affirmative, it was also suggested to introduce a new "General" category alongside the original six guiding principles to cover broader issues such as data privacy, societal considerations, and regulatory compliance, and to produce a holistic framework for trustworthy AI.

*Round 3. Feedback on the reduced set of recommendations:*

The updated version of the guideline from Round 2 was distributed to all experts for another round of feedback. This involved assessing both the adequacy and the phrasing of the recommendations. In addition, we presented the points of contention identified in the survey, encouraging experts to offer their insights on these disagreements. Examples of contentious topics included the recommendation to perform multi-centre versus local clinical evaluation, and the necessity (or not) to systematically evaluate the AI tools against adversarial attacks.

The feedback received from the experts played a crucial role in resolving several contentious issues, particularly through the refinement of the recommendations' wording. Moreover, we broadened the scope of these formulations from “AI in medical imaging” to “AI in healthcare” more generally. As a result, this led to the expansion of the FUTURE-AI guideline to a total of 30 best practices, which included 6 new recommendations within the “General” category. Areas of disagreement that remained unresolved were carefully documented and summarised for future discussions.

*Round 4. Further feedback and rating of the recommendations:*

The updated recommendations were sent out to the experts for additional feedback, this time in written form, to assess each recommendation’s clarity, feasibility, and relevance. This phase allowed for more precise phrasing of the recommendations. As an example, the original recommendation to train AI models with “diverse, heterogeneous data” was refined by using the term “representative data”, as many experts argued that representative data more effectively captures the essential characteristics of the populations, while the term heterogeneous is more ambiguous.

Furthermore, we implemented a system to rate each best practice depending on the specific needs and goals of each AI project. A key focus was to make a distinction between healthcare AI tools at the research or proof-of-concept stage and those intended for clinical deployment, as they require different levels of compliance. Healthcare AI tools in the research or proof-of-concept stage are typically in their experimental phase and require some flexibility as their capabilities are being explored and fine-tuned. In contrast, AI tools intended for clinical deployment will interact directly with patient care and, hence should need higher standards of compliance to ensure they are ethical, safe and effective. Hence, at this point of the process, the consortium members were requested to assess all the recommendations separately for both proof-of-concept and deployable AI tools and categorise them as either "recommended" or "highly recommended".

*Round 5. Feedback on the manuscript:*

At this stage, with a well-developed set of 30 recommendations, the first and last authors of the study drafted the first version of the FUTURE-AI manuscript. The draft manuscript was circulated among the experts, initiating a series of iterative feedback sessions to ensure that the FUTURE-AI guideline was articulated with precision and clarity. This process enabled incorporation of diverse perspectives, from clinical, technical, and non-technical experts, hence making the manuscript more reader-friendly and accessible to a broad audience. Experts were also able to suggest additional resources or references to substantiate the recommendations further. At this stage, examples of methods were integrated to the manuscript where relevant, aiming to demonstrate the practical implementation of the best practices in real-world scenarios.

*Round 6. New "External" feedback:*

In Round 6 we invited additional experts (*n*=44) who had not participated in the initial stages of the study to provide independent feedback. This group was carefully selected to ensure a more diverse representation across the experts (*e.g.* patient advocates, social scientists, regulatory experts), as well as wider geographic diversity (especially across Africa, Latin America, and Asia).

These experts were requested to provide written feedback and express their opinion on each recommendation using a voting system (*i.e.* Agree, Disagree, Neutral, Did not understand, No opinion). This stage was especially helpful in pinpointing any remaining areas of ambiguity or contention that required further discussions, as well as in identifying the formulations that needed refinement to ensure the entire guideline is clear and accessible to a diverse audience within the medical AI community.

*Round 7. Online consensus meetings:*

Based on the feedback from previous rounds, we identified a few topics that continued to evoke a degree of contention among experts, particularly concerning the exact wording of certain recommendations. Hence, we convened four online meetings in June 2023 specifically aimed at deepening the discussions around the remaining contentious areas and reaching a final consensus on both the recommendations and their formulations.

These discussions resolved outstanding issues such as the recommendation to systematically validate AI tools against adversarial attacks, which was considered by many experts as a cybersecurity concern, or the recommendation that the clinical evaluations should be conducted by third parties, which was deemed impractical at scale, especially in resource-limited settings.

As a result of these consensus meetings, the final list of FUTURE-AI recommendations was established, and their formulations were completed as detailed in Table 2.

*Final consensus vote:*

The very last step of the process involved a final vote on the derived recommendations, which took place through an online survey. At this stage, the final consortium consisted of 117 experts as more replied to the above recruitments. By the end of this process, all the recommendations were approved with less than 5% disagreement among all FUTURE-AI members.

## FUTURE-AI GUIDELINE

In this section, we provide definitions and justifications for each of the six guiding principles and give an overview of the FUTURE-AI recommendations. Table 2 provides a summary of the recommendations, together with the proposed level of compliance (*i.e.* recommended vs. highly recommended). Note that a glossary of the main terms used in this paper is provided in Supplementary Table 1 in the Appendix, while the main stakeholders of relevance to the FUTURE-AI framework are listed in Supplementary Table 2 in the Appendix.

*Table 2 – List of the FUTURE-AI recommendations, together with the expected compliance for both research (Res.) and deployable (Dep.) AI tools (+: Recommended, ++: Highly recommended).*

| | | **Recommendations** | Res. | Dep. |
|---|---|---|---|---|
| F | 1 | Define any potential sources of bias from an early stage | ++ | ++ |
| | 2 | Collect information on individuals' and data attributes | + | + |
| | 3 | Evaluate potential biases and, when needed, bias correction measures | + | ++ |
| | 1 | Define intended clinical settings and cross-setting variations | ++ | ++ |
| U | 2 | Use community-defined standards (*e.g.* clinical definitions, technical standards) | + | + |
| | 3 | Evaluate using external datasets and/or multiple sites | ++ | ++ |
| | 4 | Evaluate and demonstrate local clinical validity | + | ++ |
| T | 1 | Implement a risk management process throughout the AI lifecycle | + | ++ |
| | 2 | Provide documentation (*e.g.* technical, clinical) | ++ | ++ |
| | 3 | Define mechanisms for quality control of the AI inputs and outputs | + | ++ |
| | 4 | Implement a system for periodic auditing and updating | + | ++ |
| | 5 | Implement a logging system for usage recording | + | ++ |
| | 6 | Establish mechanisms for AI governance | + | ++ |
| U | 1 | Define intended use and user requirements from an early stage | ++ | ++ |
| | 2 | Establish mechanisms for human-AI interactions and oversight | + | ++ |
| | 3 | Provide training materials and activities (*e.g.* tutorials, hands-on sessions) | + | ++ |
| | 4 | Evaluate user experience and acceptance with independent end-users | + | ++ |
| | 5 | Evaluate clinical utility and safety (*e.g.* effectiveness, harm, cost-benefit) | + | ++ |
| R | 1 | Define sources of data variation from an early stage | ++ | ++ |
| | 2 | Train with representative real-world data | ++ | ++ |
| | 3 | Evaluate and optimise robustness against real-world variations | ++ | ++ |
| E | 1 | Define the need and requirements for explainability with end-users | ++ | ++ |
| | 2 | Evaluate explainability with end-users (*e.g.* correctness, impact on users) | + | + |
| General | 1 | Engage inter-disciplinary stakeholders throughout the AI lifecycle | ++ | ++ |
| | 2 | Implement measures for data privacy and security | ++ | ++ |
| | 3 | Implement measures to address identified AI risks | ++ | ++ |
| | 4 | Define adequate evaluation plan (*e.g.* datasets, metrics, reference methods) | ++ | ++ |
| | 5 | Identify and comply with applicable AI regulatory requirements | + | ++ |
| | 6 | Investigate and address application-specific ethical issues | + | ++ |
| | 7 | Investigate and address social and societal issues | + | + |

## Fairness

The Fairness principle states that AI tools in healthcare should maintain the same performance across individuals and groups of individuals (including under-represented and disadvantaged groups). AI-driven medical care should be provided equally for all citizens. Biases in healthcare AI can be due to differences in the attributes of the individuals (*e.g.* sex, gender, age, ethnicity, socioeconomic status, medical conditions) or the data (*e.g.* acquisition site, machines, operators, annotators). Fair AI tools should be developed such that potential AI biases are minimised as much as possible or identified and reported.

To this end, three recommendations for Fairness are defined in the FUTURE-AI framework:

*Fairness 1. Define sources of bias:*

Bias in healthcare AI is application-specific (27). At the design phase, the development team should identify possible types and sources of bias for their AI tool (28). These may include group attributes (*e.g.* sex, gender, age, ethnicity, socioeconomic, geography), the medical profiles of the individuals (*e.g.* with comorbidities or disability), as well as human and technical biases during data acquisition, labelling, data curation, or the selection of the input features.

*Fairness 2. Collect information on individual and data attributes:*

To identify biases and apply measures for increased fairness, relevant attributes of the individuals, such as sex, gender, age, ethnicity, risk factors, comorbidities or disabilities, should be collected. This should be subject to informed consent and approval by ethics committees to ensure an appropriate balance between the benefits of non-discrimination and the risks of re-identification. Measuring similarity of medical profiles should be also included to verify equal treatment (*e.g.* risk factors, comorbidities, biomarkers, anatomical properties (29)). Furthermore, relevant information about the datasets, such as the centres where they were acquired, the machine used, the pre-processing and annotation processes, should be systematically collected, to address technical and human biases.

*Fairness 3. Evaluate fairness:*

When possible, *i.e.* the individuals' and data attributes are available, bias detection methods should be applied by using fairness metrics such as True Positive Rates, Statistical Parity, Group Fairness, and Equalised Odds (30,31). To correct for any identified biases, mitigation measures should be tested such as data re-sampling, bias-free representations, and equalised odds post-processing (32–36) to verify their impact on both the tool's fairness and the model's accuracy. Importantly, any remaining bias should be documented and reported to inform the end-users and citizens (see Traceability 2).

**Universality**

The Universality principle states that a healthcare AI tool should be generalisable outside the controlled environment where it was built. Specifically, the AI tool should be able to generalise to new patients and new users (*e.g.* new clinicians), and when applicable, to new clinical sites. Depending on the intended radius of application, healthcare AI tools should be as interoperable and as transferable as possible, so they can benefit citizens and clinicians at scale.

To this end, four recommendations for Universality are defined in the FUTURE-AI framework:

*Universality 1. Define clinical settings:*

At the design phase, the development team should specify the clinical settings in which the AI tool will be applied (*e.g.* primary healthcare centres, hospitals, home care, low vs. high-resource settings, one or multiple countries), and anticipate potential obstacles to universality (*e.g.* differences in end-users, clinical definitions, medical equipment or IT infrastructures across settings).

*Universality 2. Use existing standards:*

To ensure the quality and interoperability of the AI tool, it should be developed based on existing community-defined standards. These may include clinical definitions of diseases by medical societies, medical ontologies (*e.g.* SNOMED CT (37)), data models (*e.g.* OMOP (38)), interface standards (*e.g.* DICOM, FHIR HL7), data annotation protocols, evaluation criteria (19), and technical standards (*e.g.* IEEE (39) or ISO (40)).

*Universality 3. Evaluate using external data:*

To assess generalisability, technical validation of the AI tools should be performed with external datasets that are distinct from those used for model training (41). These may include reference or benchmarking datasets which are representative for the task in question (*i.e.* approximating the expected real-world variations). Except for AI tools intended for single centres, the clinical evaluation studies should be performed at multiple sites to assess performance and interoperability across clinical workflows (42). If the tool's generalisability is limited, mitigation measures (*e.g.* transfer learning or domain adaptation) should be applied and tested.

*Universality 4. Evaluate local clinical validity:*

Clinical settings vary in many aspects, such as populations, equipment, clinical workflows, and end-users. Hence to ensure trust at each site, the AI tools should be evaluated for their local clinical validity (15). In particular, the AI tool should fit the local clinical workflows and perform well on the local populations. If the performance is decreased when evaluated locally, re-calibration of the AI model should be performed and tested (*e.g.* through model fine-tuning).

**Traceability**

The Traceability principle states that medical AI tools should be developed together with mechanisms for documenting and monitoring the complete trajectory of the AI tool, from development and validation to deployment and usage. This will increase transparency and accountability by providing detailed and continuous information on the AI tools during their lifetime to clinicians, healthcare organisations, citizens and patients, AI developers and relevant authorities. AI traceability will also enable continuous auditing of AI models (43), identify risks and limitations, and update the AI models when needed.

*Traceability 1. Implement risk management:*

Throughout the AI tool's lifecycle, the development team shall analyse potential risks, assess each risk's likelihood, effects and risk-benefit balance, define risk mitigation measures, monitor the risks and mitigations continuously, and maintain a risk management file. The risks may include those explicitly covered by the FUTURE-AI guiding principles (*e.g.* bias, harm, data breach), but also application-specific risks. Other risks to consider include human factors that may lead to misuse of the AI tool (*e.g.* not following the instructions, receiving insufficient training), application of the AI tool to individuals who are not within the target population, use of the tool by others than the target end-users (*e.g.* technician instead of physician), hardware failure, incorrect data annotations or input values, and adversarial attacks. Mitigation measures may include warnings to the users, system shutdown, re-processing of the input data, the acquisition of new input data, or the use of an alternative procedure or human judgement only. Monitoring and reassessment of risk may involve the use of various feedback channels, such as customer feedback and complaints, as well as logged real-world performance and issues (see Traceability 5).

*Traceability 2. Provide documentation:*

To increase transparency, traceability, and accountability, adequate documentation should be created and maintained for the AI tool (44), which may include (i) an AI information leaflet to inform citizens and healthcare professionals about the tool's intended use, risks (*e.g.* biases) and instructions for use; (ii) a technical document to inform AI developers, health organisations and regulators about the AI model's properties (*e.g.* hyperparameters), training and testing data, evaluation criteria and results, biases and other limitations, and periodic audits and updates (45–47); (iii) a publication based on existing AI reporting standards (11,13,48), and (iv) a risk management file (see Traceability 1).

*Traceability 3. Implement continuous quality control:*

The AI tool should be developed and deployed with mechanisms for continuous monitoring and quality control of the AI inputs and outputs (43), such as to identify missing or out-of-range input variables, inconsistent data formats or units, incorrect annotations or data pre-processing, and erroneous or implausible AI outputs. For quality control of the AI decisions, uncertainty estimates should be provided (and calibrated (49)) to inform the end-users on the degree of confidence in the results (50).

*Traceability 4. Implement periodic auditing and updating:*

The AI tool should be developed and deployed with a configurable system for periodic auditing (43), which should define the datasets and timelines for periodic evaluations (*e.g.* every year). The periodic auditing should enable the identification of data or concept drifts, newly occurring biases, performance degradation or changes in the decision making of the end-users (51). Accordingly, necessary updates to the AI models or AI tools should be applied (52).

*Traceability 5. Implement AI logging:*

To increase traceability and accountability, an AI logging system should be implemented to trace the user's main actions in a privacy-preserving manner, specify the data that is accessed and used, record the AI predictions and clinical decisions, and log any encountered issues. Time-series statistics and visualisations should be used to inspect the usage of the AI tool over time.

*Traceability 6. Implement AI governance:*

After deployment, the governance of the AI tool should be specified. In particular, the roles of risk management, periodic auditing, maintenance, and supervision should be assigned, such as to IT teams or healthcare administrators. Furthermore, responsibilities for AI-related errors should be clearly specified among clinicians, healthcare centres, AI developers, and manufacturers. Accountability mechanisms should be established, incorporating both individual and collective liability, alongside compensation and support structures for patients impacted by AI errors.

**Usability**

The Usability principle states that the end-users should be able to use an AI tool to achieve a clinical goal efficiently and safely in their real-world environment. On one hand, this means that end-users should be able to use the AI tool's functionalities and interfaces easily and with minimal errors. On the other hand, the AI tool should be clinically useful and safe, *e.g.* improve the clinicians' productivity and/or lead to better health outcomes for the patients and avoid harm.

To this end, five recommendations for Usability are defined in the FUTURE-AI framework:

*Usability 1. Define user requirements:*

The AI developers should engage clinical experts, end-users (*e.g.* patients, physicians) and other relevant stakeholders (*e.g.* data managers, administrators) from an early stage to compile information on the AI tool's intended use and end-user requirements (*e.g.* human-AI interfaces), as well as on human factors that may impact the usage of the AI tool (53) (*e.g.* digital literacy level, age group, ergonomics, automation bias).

*Usability 2. Define human-AI interactions and oversight:*

Based on the user requirements, the AI developers should implement interfaces to enable end-users to effectively utilise the AI model, annotate the input data in a standardised manner, and verify the AI inputs and results. Given the high-stakes nature of medical AI, human oversight is essential and increasingly required by policy makers and regulators (15,54). Human-in-the-loop mechanisms should be designed and implemented to perform specific quality checks (*e.g.* to flag biases, errors or implausible explanations), and to overrule the AI predictions when necessary.

*Usability 3. Provide training:*

To facilitate best usage of the AI tool, minimise errors and harm, and increase AI literacy, the developers should provide training materials (*e.g.* tutorials, manuals, examples) and/or training activities (*e.g.* hands-on sessions) in an accessible format and language, taking into account the diversity of end-users (*e.g.* specialists, nurses, technicians, citizens or administrators).

*Usability 4. Evaluate clinical usability:*

To facilitate adoption, the usability of the AI tool within the local clinical workflows should be evaluated in real-world setting with representative and diverse end-users (*e.g.* with respect to sex, gender, age, clinical role, digital proficiency, and disability). The usability tests should gather evidence on the user's satisfaction, performance and productivity, and assess human factors that may impact the usage of the AI tool (53) (*e.g.* confidence, learnability, automation bias).

*Usability 5. Evaluate clinical utility:*

The AI tool should be evaluated for its clinical utility and safety. The clinical evaluations of the AI tool should show benefits for the patient (*e.g.* earlier diagnosis, better outcomes), for the clinician (*e.g.* increased productivity, improved care), and/or for the healthcare organisation (*e.g.* reduced costs, optimised workflows), when compared to the current standard of care. Additionally, it is important to show that the AI tool is safe and does not cause harm to individuals (or specific groups), such as through a randomised clinical trial (RCT) (55).

**Robustness**

The Robustness principle refers to the ability of a medical AI tool to maintain its performance and accuracy under expected or unexpected variations in the input data. Existing research has shown that even small, imperceptible variations in the input data may lead AI models into incorrect decisions (56). Biomedical and health data can be subject to significant variations in the real world (both expected and unexpected), which can affect the performance of AI tools. Hence, it is important that healthcare AI tools are designed and developed to be robust against real-world variations, and evaluated and optimised accordingly.

To this end, three recommendations for Robustness are defined in the FUTURE-AI framework:

*Robustness 1. Define sources of data variations:*

At the design phase, the development team should first define robustness requirements for the AI tool in question, by making an inventory of the sources of variation that may impact the AI tool's robustness in the real world. These may include differences in equipment, technical fault of a machine, data heterogeneities during data acquisition or annotation, and/or adversarial attacks (56).

*Robustness 2. Train with representative data:*

Clinicians, citizens and other stakeholders are more likely to trust the AI tool if it is trained on data that adequately represents the variations encountered in real-world clinical practice (57). Hence, the training datasets should be carefully selected, analysed and enriched according to the sources of variation identified at the design phase (see Robustness 1).

*Robustness 3. Evaluate robustness:*

Evaluation studies should be implemented to evaluate the AI tool's robustness (*e.g.* stress tests, repeatability tests (58)), under conditions that reflect the variations of real-world clinical practice. These may include data, equipment, technician, clinician, patient and centre related variations. Depending on the results, mitigation measures should be implemented and tested to optimise the robustness of the AI model, such as regularisation (59), data augmentation (60), data harmonisation (61), or domain adaptation (62).

**Explainability**

The Explainability principle states that medical AI tools should provide clinically meaningful information about the logic behind the AI decisions. While medicine is a high-stake discipline that requires transparency, reliability and accountability, machine learning techniques often produce complex models which are black box in nature. Explainability is considered desirable from a technological, medical, ethical, legal as well as patient perspective (63). It enables end-users to interpret the AI model and outputs, understand the capacities and limitations of the AI tool, and intervene when necessary, such as to decide to use it or not. However, explainability is a complex task which has challenges that need to be carefully addressed during AI development and evaluation to ensure that AI explanations are clinically meaningful and beneficial to the end-users (64).

Two recommendations for Explainability are defined in the FUTURE-AI framework:

*Explainability 1. Define explainability needs:*

At the design phase, it should be established with end-users and domain experts if explainability is required for the AI tool. In so, the specific requirements for explainability should be defined with representative experts and end-users, including (i) the goal of the explanations (*e.g.* global description of the model's behaviour vs. local explanation of each AI decision), (ii) the most suitable approach for AI explainability (65), and (iii) the potential limitations to anticipate and monitor (*e.g.* over-reliance of the end-users on the AI decision (64)).

*Explainability 2. Evaluate explainability:*

The explainable AI methods should be evaluated, first quantitatively by using computational methods to assess the correctness of the explanations (66,67), then qualitatively with end-users to assess their impact on user satisfaction, confidence and clinical performance (68). The evaluations should also identify any limitations of the AI explanations (*e.g.* they are clinically incoherent (69) or sensitive to noise or adversarial attacks (70), they unreasonably increase the confidence in the AI-generated results (71)).

## General recommendations

Finally, seven general recommendations are defined in the FUTURE-AI framework, which apply across all principles of trustworthy AI in healthcare:

*General 1. Engage stakeholders continuously:*

Throughout the AI tool's lifecycle, the AI developers should continuously engage with inter-disciplinary stakeholders, such as healthcare professionals, citizens, patient representatives, expert ethicists, data managers and legal experts. This interaction will facilitate the understanding and anticipation of the needs, obstacles and pathways towards acceptance and adoption. Methods to engage stakeholders may include working groups, advisory boards, one-to-one interviews, co-creation meetings and surveys.

*General 2. Ensure data protection:*

Adequate measures to ensure data privacy and security should be put in place throughout the AI lifecycle. These may include privacy-enhancing techniques (*e.g.* differential privacy, encryption), data protection impact assessment and appropriate data governance after deployment (*e.g.* logging system for data access, see Traceability 5). If de-identification is implemented (*e.g.* pseudonymisation, k-anonymity), the balance between the health benefits for citizens and the risks for re-identification should be carefully assessed and considered. Furthermore, the manufacturers and deployers should implement and regularly evaluate measures for protecting the AI tool against malicious attacks, such as by using system-level cybersecurity solutions or application-specific defence mechanisms (*e.g.* attack detection or mitigation) (72).

*General 3. Implement measures to address AI risks:*

At the development stage, the development team should define an AI modelling plan that is aligned with the application-specific requirements. After implementing and testing a baseline AI model, the AI modelling plan should include mitigation measures to address the challenges and risks identified at the design stage (see Fairness 1 to Explainability 1). These may include measures to enhance robustness to real-world variations (*e.g.* regularisation, data augmentation, data harmonisation, domain adaptation), ensure generalisability across settings (*e.g.* transfer learning, knowledge distillation), and correct for biases across subgroups (*e.g.* data re-sampling, bias-free representation, equalised odds post-processing).

*General 4. Define an adequate AI evaluation plan:*

To increase trust and adoption, an appropriate evaluation plan should be defined, including test data, metrics and reference methods. First, adequate test data should be selected to assess each dimension of trustworthy AI. In particular, the test data should be well separated from the training to prevent data leakage (73). Furthermore, adequate evaluation metrics should be carefully selected, taking into account their benefits and potential flaws (74). Finally, benchmarking with respect to reference AI tools or standard practice should be performed to enable comparative assessment of model performance.

*General 5. Comply with AI regulations:*

The development team should identify the applicable AI regulations, which vary by jurisdiction and over time. For example, in the EU, the recent AI Act classifies all AI tools in healthcare as high risk, hence they must comply with safety, transparency and quality obligations and undergo conformity assessments. Identifying the applicable regulations at an early stage enables to anticipate regulatory obligations based on the AI tool's intended classification and risks.

*General 6. Investigate application-specific ethical issues:*

In addition to the well-known ethical issues that arise in medical AI (*e.g.* privacy, transparency, equity, autonomy), AI developers, domain specialists and professional ethicists should identify, discuss and address all application-specific ethical, social and societal issues as an integral part of the development and deployment of the AI tool (75).

*General 7. Investigate social and societal issues:*

In addition to clinical, technical, legal and ethical implications, a healthcare AI tool may have specific social and societal issues. These will need to be considered and addressed to ensure a positive impact for the AI tool on citizens and society. Relevant issues may include the impact of the AI tool on the working conditions and power relations, on the new skills (or deskilling) of the healthcare professionals and citizens (76), and on future interactions between citizens, health professionals and social careers. Furthermore, for environmental sustainability, AI developers should consider strategies to reduce the carbon footprint of the AI tool (77).

## OPERATIONALISATION OF FUTURE-AI

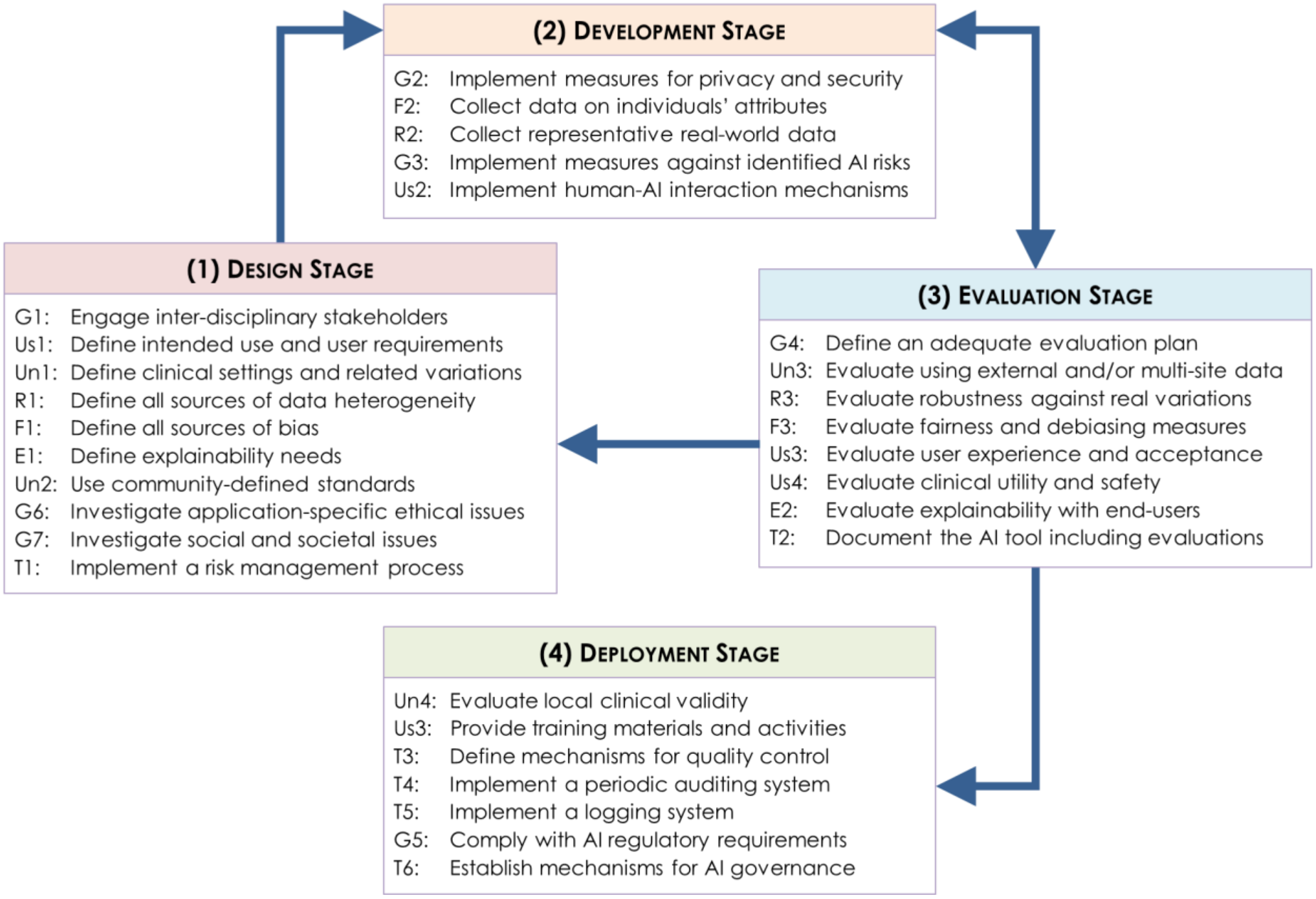

*Figure 3 – Embedding the FUTURE-AI best practices into an agile process throughout the AI lifecycle.*

To enable the implementation of the FUTURE-AI framework in practice, we provide a step-by-step guidance by embedding the recommended best practices in a chronological order across the key stages of an AI tool's lifecycle as depicted in Figure 3 and as follows:

- The design phase is initiated with a human-centred, risk-aware strategy by engaging all relevant stakeholders and conducting a comprehensive analysis of clinical, technical, ethical, and social requirements, leading to both a list of specifications and a list of risks to monitor (*e.g.* potential biases, lack of robustness, generalisability, and transparency).
- Accordingly, the development phase prioritises the collection of representative datasets for effective training and testing, ensuring they reflect variations across the intended settings, equipment, protocols, and populations as identified previously. Furthermore, an adequate AI development plan is defined and implemented given the identified requirements and risks, including mitigation strategies and human-centred mechanisms to meet the initial design's functional and ethical requirements.

- Subsequently, the validation phase comprehensively examines all dimensions of trustworthy AI, including system performance but also robustness, fairness, generalisability, and explainability, and concludes with the generation of all necessary documentation.
- Finally, the deployment phase is dedicated to ensuring local validity, providing training, implementing monitoring mechanisms, and ensuring regulatory compliance for adoption in real-world healthcare practice.

In this section, we provide a detailed list of practical steps for each recommendation, accompanied by specific examples of approaches and methods that can be applied to operationalise each step towards trustworthy AI, as shown in Table 3. This approach offers easy-to-use, step-by-step guidance for all end-users of the FUTURE-AI framework when designing, developing, validating and deploying new AI tools for healthcare.

*Table 3 – Practical steps and examples for implementing the FUTURE-AI recommendations.*

| Recommendations | Operations | Examples |
|---|---|---|
| | **- Design Stage -** | |
| Engage inter-disciplinary stakeholders (General 1) | Identify all relevant stakeholders | Patients, GPs, nurses, ethicists, data managers (78,79) |
| | Provide information on the AI tool and AI | Educational seminars, training materials, webinars (80) |
| | Set up communication channels with stakeholders | Regular group meetings, one-to-one interviews, virtual platform (81) |
| | Organise co-creation consensus meetings | One-day co-creation workshop with *n*=15 multi-disciplinary stakeholders (82) |
| | Use qualitative methods to gather feedback | Online surveys, focus groups, narrative interviews (83) |
| Define intended use and user requirements (Usability 1) | Define the clinical need and AI tool's goal | Risk prediction, disease detection, image quantification |
| | Define the AI tool's end-users | Patients, cardiologists, radiologists, nurses |
| | Define the AI model's inputs | Symptoms, heart rate, blood pressure, ECG, image scan, genetic test |
| | Define the AI tool's functionalities and interfaces | Data upload, AI prediction, AI explainability, uncertainty estimation (84) |
| | Define requirements for human oversight | Visual quality control, manual corrections (85,86) |
| | Adjust user requirements for all end-user subgroups | According to role, age group, digital literacy level (87) |
| Define intended clinical settings and cross-setting variations (Universality 1) | Define the AI tool's healthcare setting(s) | Primary care, hospital, remote care facility, home care |
| | Define the resources needed at each setting | Personnel (experience, digital literacy), medical equipment (*e.g.* > 1.5T MRI scanner), IT infrastructure |
| | Specify if the AI tool is intended for high-end and/or low-resource settings | Facilities with MRI scanners > 1.5T vs. low-field MRIs (*e.g.* 0.5T), high-end vs. low-cost portable ultrasound (88,89) |
| | Identify all cross-settings variations | Data formats, medical equipment, data protocols, IT infrastructure (90) |
| Define sources of data heterogeneity (Robustness 1) | Engage relevant stakeholders to assess data heterogeneity | Clinicians, technicians, data managers, IT managers, radiologists, device vendors |
| | Identify equipment-related data variations | Differences in medical devices, manufacturers, calibrations, machine ranges (from low-cost to high-end) (91) |
| | Identify protocol-related data variations | Differences in image sequences, data acquisition protocols (92), data annotation methods, sampling rates, pre-processing standards |
| | Identify operator-related data variations | Different in experience and proficiency, operator fatigue, subjective judgment, technique variability |
| | Identify sources of artifacts and noises | Image noise, motion artifacts, signal dropout, sensor malfunction |
| | Identify context-specific data variations | Lower data quality acquisition in emergency units, during high patient volume times |
| Define any potential sources of bias (Fairness 1) | Engage relevant stakeholders to define the sources of bias | Patients, clinicians, epidemiologists, ethicists, social carers (93,94) |
| | Define standard attributes that may impact the AI tool's fairness | Sex, age, socioeconomic status (95) |
| | Identify application-specific sources of bias beyond standard attributes | Skin colour for skin cancer detection (96,97), breast density for breast cancer detection (29) |
| | Identify all possible human biases | Data labelling, data curation (95) |
| Define the need and requirements for explainability with end-users (Explainability 1) | Engage end-users to define explainability requirements | Clinicians, technicians, patients (98) |
| | Specify if explainability is necessary | Not necessary for AI-enabled image segmentation part, critical for AI-enabled diagnosis |
| | Specify the objectives of AI explainability (if it is needed) | Understanding AI model, aiding diagnostic reasoning, justifying treatment recommendations (99) |
| | Define suitable explainability approaches | Visual explanations, feature importance, counterfactuals (100) |
| | Adjust the design of the AI explanations for all end-user subgroup | Heatmaps for clinicians, feature importance for patients (101,102) |
| | Consult ethicists on ethical considerations | Ethicists specialised in medical AI and/or in the application domain (*e.g.* paediatrics) (103) |
| Investigate ethical issues (General 6) | Assess if the AI tool's design is aligned with relevant ethical values | Right to autonomy, information, consent, confidentiality, equity (103) |
| | Identify application-specific ethical issues | Ethical risks for a paediatric AI tool (*e.g.* emotional impact on children) (104,105). |
| | Comply with local ethical AI frameworks | AI ethical guidelines from Europe (106), United Kingdom (107,108), United Sates (109), Canada (110), China (111,112), India (113), Japan (114,115), Australia (116), etc. |
| Investigate social and societal issues (General 7) | Investigate AI tool's social and societal impacts | Workforce displacement, worsened working conditions and relations, deskilling (76), dehumanisation of care, reduced health literacy, increased carbon footprint (117), negative public perception (103,118) |
| | Define mitigations to enhance the AI tool's social impact | Interfaces for physician-patient communication, workforce training, educational programs, energy-efficient computing practices, public engagement initiatives |
| Use community-defined standards (Universality 2) | Use a standard definition for the clinical task | Definition of heart failure by the American Academy of Cardiology (119) |
| | Use a standard method for data labelling | BI-RADS for breast imaging (120) |
| | Use a standard ontology for the AI inputs | DICOM for imaging data (121), SNOMED for clinical data (37) |
| | Adopt technical standards | IEEE 2801-2022 for medical AI software (39) |
| | Use standard evaluation criteria | See (19) for medical imaging applications, (30,31) for fairness evaluation. |
| Implement a risk management process (Traceability 1) | Identify all possible clinical, technical, ethical and societal risks | Bias against under-represented subgroups, limited generalisability to low-resource facilities, data drift, lack of acceptance by end-users, sensitivity to noisy inputs (122) |
| | Identify all possible operational risks | Misuse of the AI tool (due to insufficient training or not following the instructions), application of the AI tool outside of the target population (*e.g.* individuals with implants), use of the tool by others than the target end-users (*e.g.* technician instead of physician), hardware failure, incorrect data annotations, adversarial attacks (72,123) |
| | Assess the likelihood each risk | Very likely, likely, possible, rare |
| | Assess the consequences of each risk | Patient harm, discrimination, lack of transparency, loss of autonomy, patient re-identification (124) |
| | Prioritise all the risks depending on their likelihood and consequences | Risk of bias (if no personal attributes are included in the model) vs. risk of patient re-identification (if personal attributes are collected) |
| | Define mitigation measures to be applied during AI development | Data enhancement, data augmentation (125), bias correction techniques, domain adaptation (62), transfer learning (126), continuous learning (127) |
| | Define mitigation measures to be applied post deployment | Warnings to the users, system shutdown, re-processing of the input data, the acquisition of new input data, the use of an alternative procedure or human judgement only. |
| | Set up a mechanism to monitor and manage risks over time | Periodic risk assessment every six months |
| | Create a comprehensive risk management file | Including all risks, their likelihood and consequences, risk mitigation measures, risk monitoring strategy |
| **- Development Stage -** | | |
| Collect representative training dataset (Robustness 2) | Collect training data that reflect the demographic variations | According to age, sex, ethnic, socioeconomics. |
| | Collect training data that reflect the clinical variations | Disease subgroups, treatment protocols, clinical outcomes, rare cases. |
| | Collect training data that reflect variations in real-world practice | Data acquisition protocols, data annotations, medical equipment, operational variations (*e.g.* patient motion during scanning) (123) |
| | Artificially enhance the training data to mimic real-world conditions | Data augmentation (125), data synthesis (*e.g.* low-quality data, noise addition) (128), data harmonisation (129,130), data homogenisation (131) |
| Collect information on individuals' and data attributes (Fairness 2) | Request approval for collecting data on personal attributes | Sex, age, ethnicity, socioeconomic status (132) |
| | Collect information on standard attributes of the individuals (if available and allowed) | Sex, age, nationality, education (133) |
| | Include application-specific information relevant for fairness analysis | Skin colour, breast density (29), presence of implants, comorbidity (134) |
| | Estimate data distributions across subgroups | Male vs. female, across ethnic groups |
| | Collect information on data provenance | Data centres, equipment characteristics, data pre-processing, annotation processes |
| Implement measures for data privacy and security (General 2) | Implement measures to ensure data privacy and security | Data de-identification, federated learning (135–137), differential privacy, encryption (138) |
| | Implement measures against malicious attacks | Firewalls, intrusion detection systems, regular security audits (138) |
| | Adhere to applicable data protection regulations | General Data Protection Regulation (139), Health Insurance Portability and Accountability Act (140) |
| | Define suitable data governance mechanisms | Access control, logging system |
| Implement measures against identified AI risks (General 3) | Implement a baseline AI model and identify its limitations | Bias, lack of generalisability (141) |
| | Implement methods to enhance robustness to real-world variations (if needed) | Regularisation (142), data augmentation (125), data harmonisation (129), domain adaptation (62) |
| | Implement methods to enhance generalisability across settings (if needed) | Regularisation, transfer learning (143), knowledge distillation (144) |
| | Implement methods to enhance fairness across subgroups (if needed) | Data re-sampling, bias-free representation(32), equalised odds post-processing (33,34,145) |
| Establish mechanisms for human-AI interactions (Usability 2) | Implement mechanisms to standardise data pre-processing and labelling | Data pre-processing pipeline, data labelling plugin. |
| | Implement an interface for utilising the AI model | Application programming interface |
| | Implement interfaces for explainable AI | Visual explanations, heatmaps, feature importance bars (101,102) |
| | Implement mechanisms for user-centred quality control of the AI results | Visual quality control, uncertainty estimation (146) |
| | Implement mechanism for user feedback | Feedback interface (147) |
| **- Evaluation Stage -** | | |
| Define adequate evaluation plan (General 4) | Identify the dimensions of trustworthy AI to be evaluated | Robustness, clinical safety, fairness, data drifts, usability, explainability |
| | Select appropriate testing datasets | External dataset from a new hospital, public benchmarking dataset (147) |
| | Compare the AI tool against standard of care | Conventional risk predictors, visual assessment by radiologist, decision by clinician (148,149) |
| | Select adequate evaluation metrics | F1-score for classification, concordance index for survival (19), statistical parity for fairness (150). |
| Evaluate using external datasets and/or multiple sites (Universality 3) | Identify relevant public datasets | The Cancer Imaging Archive (151), the UK Biobank (152), M&Ms (153), MAMA-MIA (154), BRATS (155) |
| | Identify external private datasets | New prospective dataset from same site or from a different clinical centre (156,157) |
| | Select multiple evaluation sites | Three sites in same country, five sites in two different countries |
| | Verify that the evaluation data and sites reflect real-world variations | Variations in demographics, clinicians, equipment |
| | Confirm that no evaluation data was used during training | Yes/no |
| Evaluate fairness and bias correction measures (Fairness 3) | Select attributes and factors for fairness evaluation | Sex, age, skin colour, comorbidity |
| | Define fairness metrics and criteria | Statistical parity difference defined fairness between [-0.1 ,0.1] (31) |
| | Evaluate fairness and identify biases | Fair with respect to age, biased with respect to sex |
| | Evaluate bias mitigation measures | Training data re-sampling (158), equalised odds post-processing (33,34,145) |
| | Evaluate the impact of the mitigation measures on model performance | Data re-sampling removed sex bias but reduced model performance (159) |
| | Report identified and uncorrected biases | In the AI information leaflet and technical documentation (160) (see Traceability 2). |
| Evaluate user experience (Usability 4) | Evaluate usability with diverse end-users | According to sex, age, digital proficiency level, role, clinical profile (161,162) |
| | Evaluate user satisfaction using usability questionnaires | System usability scale (163) |
| | Evaluate user performance and productivity | Diagnosis time with and without the AI tool, image quantification time (164) |
| | Assess the training of new end-users | Average time to reach competency, training difficulties (165) |
| Evaluate clinical utility and safety (Usability 5) | Define clinical evaluation plan | Randomised control trial (RCT) (55,166), in-silico trial (167) |
| | Evaluate if the AI tool improves patient outcomes | Better risk prevention, earlier diagnosis, more personalised treatment (168) |
| | Evaluate if AI tool enhances productivity or quality of care | Enhanced patient triage, shorter waiting times, faster diagnosis, higher patient intake (168) |
| | Evaluate if AI tool results in cost savings | Reduction in diagnosis costs (169,170), reduction in over-treatment (171) |
| | Evaluate AI tool’s safety | Side effects or major adverse events in RCTs (172,173) |
| Evaluate robustness (Robustness 3) | Evaluate robustness under real-world variations | Using test-retest datasets (174,175), multi-vendor datasets (176) |
| | Evaluate robustness under simulated variations | Using simulated repeatability tests (147), synthetic noise and artefacts (*e.g.* image blurring) (177) |
| | Evaluate robustness against variations in end-users | Different technicians or annotators |
| | Evaluate mitigation measures for robustness enhancement | Regularisation (59), data augmentation (60,125), noise addition, normalisation (178), resampling, domain adaptation (62) |
| Evaluate explainability (Explainability 2) | Assess if the explanations are clinically meaningful | Reviewing by expert panels, alignment to current clinical guidelines, explanations not pointing to shortcuts (69) |
| | Assess explainability quantitatively using objective measures | Fidelity, consistency, completeness, sensitivity to noise (179–181) |
| | Assess explainability qualitatively with end-users | Using user tests or questionnaires to measure confidence and impact on clinical decision making (182,183) |
| | Evaluate if the explanations cause end-user over-confidence or over-reliance | Measure changes in clinician confidence (184,185), performance with and without AI tool (186) |
| | Evaluate if the explanations are sensitive to input data variations | Stress tests under perturbations to evaluate the stability of explanations (70,187) |
| Provide documentation (Traceability 2) | Report evaluation results in publication using AI reporting guidelines | Peer-reviewed scientific publication using TRIPOD-AI reporting guideline (13) |
| | Create technical documentation for the AI tool | AI passport (188), model cards (45) (including model hyperparameters, training and testing data, evaluations, limitations, etc |
| | Create clinical documentation for the AI tool | Guidelines for clinical use, AI information leaflet (including intended use, conditions and diseases, targeted populations, instructions, potential benefits, contra-indications) |
| | Provide a risk management file | Including identified risks, mitigation measures, monitoring measures |
| | Create user and training documentation | User manuals, training materials, troubleshooting, FAQs (See Usability 2) |
| | Identify and provide all locally required documentation | Compliance documents and certifications (see General 5) |
| **- Deployment Stage -** | | |
| Evaluate and demonstrate local clinical validity (Universality 4) | Test the AI model using local data | Data from the local clinical registry |
| | Identify factors that could impact the AI tool’s local validity | Local operators, equipment, clinical workflows, acquisition protocols |
| | Assess the AI tool’s integration within local clinical workflows | The AI tool’s interface aligns with the hospital IT system (147) or disrupts routine practice |
| | Assess the AI tool's local practical utility and identify any operational challenges | Time to operate, clinician satisfaction, disruption of existing operations (147,189) |
| | Implement adjustments for local validity | Model calibration, fine-tuning (190), transfer learning (191–193) |
| | Compare performance of AI tool to that of the local clinicians | Side-by-side comparison, in-silico trial |
| Define mechanisms for quality control of the AI inputs and outputs (Traceability 3) | Implement mechanisms to identify erroneous input data | Missing value or out-of-distribution detector (194), automated image quality assessment (69,195,196) |
| | Implement mechanisms to detect implausible AI outputs | Post-processing sanity checks, anomaly detection algorithm (197) |
| | Provide calibrated uncertainty estimates to inform on the AI tool's confidence | Calibrated uncertainty estimates per patient or data point (49,50,198) |
| | Implement a system for continuous quality monitoring | Real-time dashboard tracking data quality and performance metrics (199) |
| | Implement a feedback mechanism for users to report issues | Feedback portal enabling clinicians to report discrepancies or anomalies |
| Implement a system for periodic auditing and updating (Traceability 4) | Define a schedule for the periodic audits | Biannual or annual |
| | Define audit criteria and metrics | Accuracy, consistency, fairness, data security (147) |
| | Define datasets for the periodic audits | Newly acquired prospective dataset from the local hospital |
| | Implement mechanisms to detect data or concept drifts | Detecting shifts in input data distributions (147,189) |
| | Assign the role of auditor(s) for the AI tool | Internal auditing team, third-party company (189) |
| | Update AI tool based on audit results | Updating AI model (52), re-evaluating AI model (147), adjusting operational protocols, continuous learning (200–203) |
| | Implement reporting system from audits and subsequent updates | Automatic sharing of detailed reports to healthcare managers and clinicians |
| | Monitor impact of AI updates | Impact on system performance and user satisfaction (52) |
| Implement a logging system for usage recording (Traceability 5) | Implement a logging framework capturing all interactions | User actions, AI inputs, AI outputs, clinical decisions |
| | Define the data to be logged | Timestamp, user id, patient id (anonymised), action details, results |
| | Implement mechanisms for data capture | Software to automatically record every data and operation |
| | Implement mechanisms for data security | Encrypted log files, privacy-preserving techniques (204) |
| | Provide access to logs for auditing and troubleshooting | By defining authorised personnel, *e.g.* healthcare or IT managers |
| | Implement a mechanism for the end-users to log any issues | A user interface to enter information about operational anomalies |
| | Implement log analysis | Time-series statistics and visualisations to detect unusual activities and alert administrators |
| Provide training (Usability 3) | Create user manuals | User instructions, capabilities, limitations, troubleshooting steps, examples and case studies |
| | Develop training materials and activities | Online courses, workshops, hands-on sessions |
| | Use formats and languages accessible to intended end-users | Multiple formats (text, video, audio) and languages (English, Chinese, Swahili) |
| | Customise training to all end-user groups | Role-specific modules for specialists, nurses and patients |
| | Include training to enhance AI and health literacy | On application-specific AI concepts (*e.g.* radiomics, explainability), AI-driven clinical decision making |
| Identify and comply with applicable AI regulatory requirements (General 5) | Engage regulatory experts to investigate regulatory requirements | Regulatory consultants from intended local settings |
| | Identify specific regulations based on AI tool's intended markets | FDA's Software as a Medical Device (SaMD) in the US (205), Medical Device Regulation (MRD) and AI Act (206) in the EU |
| | Identify the specific requirements based on AI tool's purpose | De Novo classification (Class III) (207) |
| | Define a list of milestones towards regulatory compliance | MDR certification: technical verification, pivotal clinical trial, risk and quality management, post-market follow-up |
| Establish mechanisms for AI governance (Traceability 6) | Assign roles for the AI tool's governance | For periodic auditing, maintenance, supervision (*e.g.* healthcare manager) |
| | Define responsibilities for AI-related errors | Responsibilities of clinicians, healthcare centres, AI developers and manufacturers |
| | Define mechanisms for accountability | Individual vs. collective accountability/liability (23), compensations, support for patients |

## DISCUSSION

Despite the tremendous amount of research in medical AI in recent years, currently only a limited number of AI tools have made the transition to clinical practice. While many studies have demonstrated the huge potential of AI to improve healthcare, significant clinical, technical, socio-ethical and legal challenges persist.

In this paper, we presented the results of an international effort to establish a consensus guideline for developing trustworthy and deployable AI tools in healthcare. Through an iterative process that lasted 24 months, the FUTURE-AI framework was established, comprising a comprehensive and self-contained set of 30 recommendations, which covers the whole lifecycle of medical AI. By dividing the recommendations across six guiding principles, the pathways towards responsible and trustworthy AI are clearly characterised.

By the end of the process, all the recommendations were approved with less than 5% disagreement among all FUTURE-AI members. The FUTURE-AI consortium provided knowledge and expertise across a wide range of disciplines and stakeholders, resulting in consensus and wide support, both geographically and across domains. Hence, the FUTURE-AI guideline can benefit a wide range of stakeholders, as detailed in Table 2 in the Appendix.

FUTURE-AI is a risk-informed framework. It proposes to assess application-specific risks and challenges early in the process (*e.g.* risk of discrimination, lack of generalisability, data drifts over time, lack of acceptance by end-users, potential harm for patients, lack of transparency, data security vulnerabilities, ethical risks), then implement tailored measures to reduce these risks (*e.g.* collect data on individuals' attributes to assess and mitigate bias). This is also a risk-benefit balancing exercise, as the specific measures to be implemented have benefits and potential weaknesses that the developers need to assess and take into consideration. For example, collecting data on individuals' attributes may increase the risk of re-identification, but can enable to reduce the risk of bias and discrimination. Hence, in FUTURE-AI, risk management (as recommended in Traceability 1) must be a continuous and transparent process throughout the AI tool's lifecycle.

Furthermore, FUTURE-AI is an assumption-free, highly collaborative framework. It recommends to continuously engage with multi-disciplinary stakeholders to understand application-specific needs, risks and solutions (General 1). This is crucial to remove assumptions and investigate all possible risks and factors that may reduce trust in a given AI tool. For example, instead of making any assumption on possible sources of bias (*e.g.* sex or age), FUTURE-AI recommends that the developers engage with healthcare professionals, domain experts, representative citizens, and/or ethicists early in the process to investigate in depth the application-specific sources of bias, that may include factors well beyond standards attributes (*e.g.* breast density for AI applications in breast cancer).

For deployable AI tools, 26 recommendations out of 30 are rated as highly recommended (Table 2). For research and proof-of-concept AI tools, only 12 recommendations are rated as highly recommended, but we advise that researchers use as many elements as possible from the FUTURE-AI guideline to facilitate future transitions towards real-world practice.

The FUTURE-AI guideline was defined in a generic manner to ensure it can be applied across a variety of domains (*e.g.* radiology, genomics, mobile health, electronic health records). However, for many recommendations, their applicability varies across medical use cases. Hence, the first recommendation in each of the guiding principles is to identify the specificities to be addressed, such as the types of biases (Fairness 1), the clinical settings (Universality 1), or the need and approaches for explainable AI (Explainability 1).

Furthermore, we focused on developing best practices for enhancing the trustworthiness of medical AI tools, while consciously avoiding the imposition of specific techniques for the implementation of each recommendation. This flexibility acknowledges the diversity of methods for tackling challenges and mitigating risks in medical AI. For example, the recommendation to protect personal data during AI training can be implemented through data de-identification, federated learning, differential privacy or encryption, among other methods. While such examples are listed in the manuscript, the most adequate techniques for implementing each recommendation should be ultimately selected by the AI development team as a function of the application domain, clinical use case, and data characteristics, as well as the advantages and limitations of each method.

While the FUTURE-AI framework offers insights for regulating medical AI, future work is needed to incorporate these recommendations into regulatory procedures. For example, we propose mechanisms to enhance traceability and governance, such as through AI logging. However, the crucial issue of liability is yet to be addressed (*e.g.* who should perform the audits and who should be accountable for errors). Furthermore, we recommend continuous evaluation and fine-tuning of the AI models over time. However, current regulations prevent post-release modifications, as they would formally invalidate the manufacturer's initial validation. Future regulations should address the possibility of local adaptations within pre-defined acceptance criteria.

Finally, progressive development and adoption of medical AI tools will lead to new requirements, challenges, and opportunities. Aware of this reality, we propose FUTURE-AI as a dynamic, living framework. To refine the FUTURE-AI guideline and learn from other voices, we set up a dedicated webpage (www.future-ai.eu) through which we invite the community to join the FUTURE-AI network and provide feedback based on their own experience and perspective. On the website we include a FUTURE-AI self-assessment checklist, which comprises a set of questions and examples to facilitate and illustrate the use of the FUTURE-AI recommendations. Additionally, we plan to organise regular outreach events such as webinars and workshops to exchange with medical AI researchers, manufacturers, evaluators, end-users, and regulators.

## ACKNOWLEDGMENTS

This work has been supported by the European Union's Horizon 2020 under Grant Agreement No. 101034347 (OPTIMA), No. 101016775 (INTERVENE) and No. 116074 (BigData@Heart). This work received support from the European Union's Horizon Europe under Grant Agreement No. 101057699 (RadioVal) and No. 101057849 (DataTools4Heart). This work received support from the European Research Council under Grant Agreement No. 757173 (MIRA), No. 884622 (Deep4MI), No. 101002198 (NEURAL SPICING) and No. 866504 (CANCER-RADIOMICS). This work was partially supported from the Royal Academy of Engineering, Hospital Clinic

Barcelona, Malaria No More, Carnegie Cooperation New York, Human frontier science program, Natural Sciences and Engineering Research Council of Canada (NSERC), the Australian National Health and Medical Research Council Ideas under Grant No. 1181960, United States Department of Defense W81XWH2010747-P1, 3IA Côte d'Azur Investments in the Future project managed by the National Research Agency (ANR-19-P3IA-0002), InTouchAI.eu, IITP grant funded by the Korean government (No.2020-0-00594), A*STAR Career Development Award (project No. C210112057) from the Agency for Science, Technology and Research (A*STAR), National Institute for Health and Care Research Barts Biomedical Research Centre, Centre National de la Recherche Scientifique (CNRS), MPaCT-Data. Infraestructura de Medicina de Precisión asociada a la Ciencia y la Tecnología. (Exp. IMP/00019) funded by Instituto de Salud Carlos III and the Fondo Europeo de Desarrollo Regional (FEDER, "Una manera de hacer Europa"), Ministry of Science, Technology and Innovation of Colombia project code 110192092354, Gordon and Betty Moore Foundation, Google Award for Inclusion Research, Fraunhofer Heinrich Hertz Institute, USA National Institutes of Health, National Council for Scientific and Technological Development (CNPq), European Heart Network, NIBIB/University of Chicago (MIDRC), Hong Kong Research Grants Council Theme-based Research Scheme (TRS) project T45-401/22-N, Young Researcher Project (19PEJC09-03) funded by the Ministry of High Education of Tunisia, Juan de la Cierva with reference number FJC2021-047659-I, Nepal Applied Mathematics and Informatics Institute for Research (NAAMII), Fogarty International Center of the National Institutes of Health under Award No. 5U2RTW012131-02, Universidad Galileo, Natural Science Foundation of China under Grant 62271465, Israel Science Foundation, National Institutes of Health (NIH), Dutch Cancer Society (KWF Kankerbestrijding) under project number 14449, Netherlands Organisation for Scientic Research (NWO) VICI project VI.C.182.042., National Center for Artificial Intelligence CENIA FB210017; Basal ANID, Google Research, Independent Research Fund Denmark (DFF, grant number 9131-00097B), Wellcome Flagship Programme (WT213038/Z/18/Z), Cancer Research UK programme grant (C49297/A27294) , the MIDRC (The Medical Imaging and Data Resource Center), made possible by the National Institute of Biomedical Imaging and Bioengineering (NIBIB) of the National Institutes of Health under contract 75N92020D00021, and the Employee European Heart Network.

## COMPETING INTERESTS

GD owns equity interest in Artrya Ltd. and provides consultancy services. JK-C receives research funding from GE, Genetech and is a consultant at Siloam Vision, Inc. GPK advises some AI startups such as Gleamer.AI, FLUIDDA BV, NanoX Vision. GPK was the founder of Quantib BV. SEP is a consultant for Circle Cardiovascular Imaging Inc., Calgary, Alberta, Canada. BG is employed by Kheiron Medical Technologies and HeartFlow. PL receives research/grant agreements from Radiomics SA and Convert Pharmaceuticals and has minority shares in Radiomics SA and Convert pharmaceuticals. PL is co-inventor of two issued patents with royalties on radiomics (PCT/NL2014/050248 and PCT/NL2014/050728) licensed to Radiomics SA; one non-patented inventions (software) licensed Radiomics SA and two non-issued, non-licensed patents on Deep Learning-Radiomics (N2024482, N2024889). ARP serves as advisor for mGeneRX in exchange for equity. JM receives royalties from GE, research grants from Siemens and is unpaid consultant for Nuance. HCW own minority shares in the company Radiomics SA. JWG serves on several radiology society AI committees. CL is a shareholder and advisor to Bunker Hill Health, GalileoCDS, Sirona Medical, Adra, and Kheiron Medical. He serves as a board member of Bunker Hill Health and a shareholder of whiterabbit.ai. He has served as a paid consultant to Sixth Street and Gilmartin Capital. His institution has received grants or gifts from Bunker Hill Health, Carestream, CARPL, Clairity, GE Healthcare, Google Cloud, IBM, Kheiron, Lambda, Lunit, Microsoft, Philips, Siemens Healthineers, Stability.ai, Subtle Medical, VinBrain, Visiana, Whiterabbit.ai, the Lowenstein Foundation, and the Gordon and Betty Moore Foundation. GSC is a statistics editor for the BMJ and a National Institute for Health and Care Research (NIHR) Senior Investigator. The views expressed in this article are those of the author(s) and not necessarily those of the NIHR, or the Department of Health and Social Care. All other authors declare no competing interests.

## FUNDING

Support for this work was partially provided by the European Union's Horizon 2020 under Grant Agreement No. 952103 (EuCanImage), No.952159 (ProCAncer-I), No.952172 (CHAIMELEON), No. 826494 (PRIMAGE) and No. 952179 (INCISIVE).

## AUTHOR CONTRIBUTIONS

KL, RO, NL, KK, GT, SA, LCA, KM, MT, SC, NP, ZS, HCW, PL and LMB conceptualised the FUTURE-AI framework and provided the first set of recommendations. All co-authors participated in the surveys and provided feedback throughout the process. KL organised four online meetings to discuss the final recommendations. AE and XPB coordinated the last consensus survey. KL and MPAS coordinated the feedback gathering process and wrote the manuscript. All authors reviewed and approved the manuscript.

***Appendix Table 1*** *– A glossary of main terms used in the FUTURE-AI guideline (ranked alphabetically).*

| Term | Definition |
|---|---|
| AI auditing | A periodic evaluation of an AI tool to assess its performance and working conditions over time, and to identify potential problems. |
| AI deployment | The process of placing a completed AI tool into a live clinical environment where it can be used for its intended purpose. |
| AI design | Early stage of an AI's production lifetime, during which specifications and plans are defined for the subsequent development of the AI tool |
| AI development | The process of training AI models and building AI-human interfaces, based on the specifications and plans from the AI design phase. |
| AI evaluation | The assessment of an AI tool's added value in its intended clinical setting. |
| AI model | A program trained using a machine learning algorithm to perform a given task based on specific input data. |
| AI monitoring | The process of tracking the behaviour of a deployed AI tool over time, to identify potential degradation in performance and implement mitigation measures such as model updating. |
| AI regulation | A set of requirements and obligations defined by public authorities, that AI developers, deployers and users must adhere to. |
| AI risk | Any negative effect that may occur when using an AI tool. |
| AI tool | A software that comprises the AI model plus a user interface that can be used by the end-users to perform a given AI-powered clinical task. |
| AI training | The process of using machine learning algorithms to build AI models that learn to perform specific tasks based on existing data samples. |
| AI updating | The process of re-training or fine-tuning the AI model after some time to improve its performance and correct identified issues. |
| AI validation | The assessment of an AI model's performance. |
| Attribute | Personal quality, trait or characteristic of an individual or group of individuals, such as sex, gender, age, ethnicity, socioeconomic status or disability. Protected attributes refer to those attributes that, by law, cannot be discriminated against (*i.e.* attributes that are protected by law). |
| Benchmarking | The practice of comparing the performance of multiple AI tools (or an AI tool against the standard practice) based on a common reference dataset and a set of predefined performance criteria and metrics. |
| Bias | Systematic, prejudiced errors by an AI tool against certain individuals or subgroups due to inadequate data or assumptions used during the training of the machine learning model. |
| Clinical safety | The capability of an AI tool to keep individuals and patients safe and not to cause them any harm. |
| Clinical setting | The environment or location where the AI tool will be used, such as a hospital, a radiology department, a primary care centre, or for home-based care. |
| Clinical utility | The capability of an AI tool to be useful in its intended clinical settings, such as to improve clinical outcomes, to increase the clinicians' productivity, or to reduce healthcare costs. |
| Concept drift | Changes in relationship between AI model inputs and outputs. |
| Data drift | Changes in the distribution of the AI model's input data over time. |
| Data quality control | The process of assessing the quality of the input data, to identify potential defects that may affect the correct functioning of the AI tool. |
| Deployable AI | AI developed with a high technology readiness level (TRL) (5-9) intended for deployment in clinical practice. |
| Ethical AI | AI that adheres to key ethical values and human rights, such as the rights to privacy, equity and autonomy. |
| Explainability | The ability of an AI tool to provide clinically meaningful information about the logic behind the AI decisions. |
| Fairness | The ability of an AI tool to treat equally individuals with similar characteristics or subgroups of individuals including under-represented groups. |
| Human oversight | A procedure or set of procedures put in place to ensure an AI tool is used under the supervision of a human (*e.g.* a clinician), who is able to overrule the AI decisions and take the final clinical decision. |
| Intended use | Clinical purpose or clinical task that the AI tool aims to realise in its intended clinical setting. |
| Logging | The process of keeping a log of events that occur while using an AI tool, such as user actions, accessed and used datasets, clinical decisions, and identified issues. |
| Proof-of-concept AI | AI developed with a low machine learning technology readiness level (ML-TRL) (1-4) to demonstrate the feasibility of a new AI method or new AI concept. |
| Real world | The clinical environment in which AI tools will be applied in practice, outside the controlled environment of research labs. |
| Responsible AI | AI that is designed, developed, evaluated, and monitored by employing an appropriate code of conduct and appropriate methods to achieve technical, clinical, ethical, and legal requirements (*e.g.* efficacy, safety, fairness, robustness, transparency). |
| Robustness | The ability of an AI tool to overcome expected or unexpected variations, such as due to noise or artefacts in the data. |
| Third-party evaluator | An independent evaluator who did not participate in any way in the design or development of the AI tool to be evaluated. |
| Traceability | The ability of an AI tool to be monitored over its complete lifecycle. |
| Trustworthy AI | AI with proven characteristics such as efficacy, safety, fairness, robustness, transparency, which enable relevant stakeholders such as citizens, clinicians, health organisations and authorities to rely on it and adopt it in real-world practice. |
| Trustworthy AI vs. Responsible AI | For trustworthy AI, the emphasis is on the characteristics of the AI tool and how they are perceived by the stakeholders of interest (*e.g.* patients, clinicians), while for responsible AI, the emphasis is on the developers, evaluators and managers of the AI tool, and the code of conduct and methods they employ to obtain trustworthy AI tools. |
| Universality | The ability of an AI tool to generalise across clinical settings. |
| Usability | The degree to which an AI tool is fit to be used by end-users in the intended clinical setting. |

***Table 2*** *– List of stakeholder groups (ranked alphabetically) that can benefit from the FUTURE-AI guideline.*

| Stakeholders | FUTURE-AI usage |
|---|---|
| AI ethicists | o To embed ethics into the development of medical AI tools. |
| AI evaluators/clinical trialists | o To perform more comprehensive, multi-faceted evaluations of medical AI tools based on the principles of trustworthy AI.<br>o To assess the trustworthiness of AI tools. |
| Citizens and patients | o To increase literacy about medical AI and trustworthy AI.<br>o To increase engagement in the production and evaluation of medical AI tools. |
| Conferences/journals | o To promote best practices and new methods for trustworthy AI among researchers reading or publishing scientific papers. |
| Data managers | ○ To support the development and deployment of medical AI tools that are compliant with data protection/governance principles. |
| Educational institutions | ○ To educate students from all disciplines (machine learning, computer science, medicine, ethics, social sciences) on the principles and approaches for trustworthy AI. |
| Funding agencies | ○ To promote new research projects that integrate best practices and new approaches for responsible AI. |
| Health organisations | ○ To guide healthcare organisations in the evaluation, deployment and monitoring of medical AI tools.<br>○ To verify the trustworthiness of AI tools. |
| Healthcare professionals | ○ To adopt the principles of trustworthy AI and best practices among the healthcare professions.<br>○ To engage clinicians in the design, development, evaluation and monitoring of medical AI tools. |
| IT managers | ○ To promote IT solutions for the deployment and monitoring of trustworthy and secure AI tools in clinical practice. |
| Legal experts | ○ To ensure compliance with applicable laws and regulations related to medical AI and data protection. |
| Manufacturers of medical AI devices | ○ To adopt best practices for responsible AI within companies.<br>○ To develop and/or commercialise new AI tools that will be accepted, certified and deployed for clinical use. |
| Public authorities | ○ To adapt existing regulations and policies on medical AI. |
| Regulatory bodies | ○ To enhance the procedures for the evaluation, certification and monitoring of AI tools as medical devices. |
| Researchers and developers in medical AI | ○ To investigate new methods according to the recommendations for trustworthy AI.<br>○ To develop proof-of-concepts that can more easily transition into deployable AI tools for clinical practice. |
| Scientific/medical societies | ○ To promote the principles of trustworthy AI and best practices among scientific and medical communities. |
| Social scientists | ○ To ensure social and societal dimensions of medical AI are considered. |
| Standardisation bodies | ○ To develop new standards that facilitate the implementation, evaluation and adoption of trustworthy AI tools in healthcare. |